\documentclass[aps,twocolumn,prl,superscriptaddress,letterpaper]{revtex4}

\usepackage{amsmath,amssymb}
\usepackage[dvips]{graphicx}
\usepackage{longtable}
\usepackage[usenames]{color}
\usepackage[normalem]{ulem}
\usepackage{bm}

\begin{document}

\newcommand{\bra}[1]{\langle #1|}
\newcommand{\ket}[1]{|#1\rangle}\newcommand{\B}[1]{\textcolor{blue}{#1}}
\newcommand{\R}[1]{\textcolor{red}{#1}}
\newcommand{\be}{\begin{equation}}
\newcommand{\ee}{\end{equation}}
\newcommand{\ba}{\begin{eqnarray}}
\newcommand{\ea}{\end{eqnarray}}

\title{Reveal non-Markovianity of open quantum systems via local operations}
\author{Huan Yang}
\affiliation{Theoretical Astrophysics 350-17, California Institute
of Technology, Pasadena, CA 91125, USA}
\author{Haixing Miao}
\affiliation{Theoretical Astrophysics 350-17, California Institute
of Technology, Pasadena, CA 91125, USA}
\author{Yanbei Chen}
\affiliation{Theoretical Astrophysics 350-17, California Institute
of Technology, Pasadena, CA 91125, USA}

\begin{abstract}
Non-Markovianity, as an important feature of general open quantum systems, is usually
difficult to quantify with limited knowledge of how the plant that we are interested in
interacts with its environment---the bath. It often happens that the reduced dynamics of
the plant attached to a non-Markovian bath becomes indistinguishable from the one with
a Markovian bath, if we left the entire system freely evolve.  Here we show that
non-Markovianity can be revealed via applying local unitary operations on the plant---they
will influence the plant evolution at later times due to memory of the bath.  
This not only provides a new criterion for non-Markovianity, but also sheds light on 
protecting and recovering quantum coherence in non-Markovian systems, which will 
be useful for quantum-information processing.
\end{abstract}
\maketitle

{\it Introduction.}---A systematic characterization of non-Markovianity in open quantum systems
has recently become an active field of both theoretical and experimental study~\cite{Breuer, Wolf, Breuer2, Rivas, Liu},
which is mostly motivated by recent quests for robust quantum-information protocols. Understanding
non-Markovianity can lead to novel strategies for protecting the quantum coherent of the plant---the
part of  quantum system that we care about, e.g., the atomic spin. Especially when we have certain
control over the plant, non-Markovianity of the bath, in principle, allows us to decouple the plant
from the bath, which is known as dynamical decoupling\,\cite{Viloa, Biercuk, Lange}.

Although non-Markovian dynamics arises ubiquitously as long as the bath has a temporal response
comparable to the dynamical time-scale of the plant, quantifying it systematically
is not straightforward for general open quantum systems. In the literature, Breuer {\it et al.}
proposed a measure based upon the evolution of the trace distance ${\rm Tr}|\hat \rho_1(t)-\hat \rho_2(t)|/2$
between two different initial quantum states of the plant $\hat \rho_1(0)$ and $\hat \rho_2(0)$\,\cite{Breuer2}:
an increase of the trace distance gives a unequivocal signature of non-Markovianity, as it indicates
that information flows from the bath back to the plant. Alternatively,  Rivas {\it et al.} proposed
introducing an ancilla to entangle but not interact with the plant: an increase in the entanglement
between the plant and the ancilla during evolution signifies the existence of non-Markovian dynamics
between the plant and the bath~\cite{Rivas}. These two important measures have been applied extensively
in studying non-Markovian quantum systems, and has been compared
theoretically\,\cite{Haikka,Chruscinski} and experimentally by Liu {\it et al.}\,\cite{Liu}.

We notice that the above-mentioned measures are focused on the reduced dynamics of the plant which contains
a limited amount of knowledge of the entire plant-bath system. It often happens that the law of time
evolution of the plant becomes identical to that of a Markovian system, even though the bath
still carries non-trivial information about the plant. Here, we propose a new criterion for
non-Markovianity, assuming that we not only have access to the time evolution of the plant's density
matrix, but can also carry out unitary operations on the plant (but not on the bath): {\it the dynamics is
non-Markovian if a local unitary operation on the plant at a given moment can influence the plant's
law of evolution at later times.} We will use this criterion to reveal non-Markovianity in systems where 
the plant follows the time-local Markovian master equation:
\be
\dot {\hat \rho}_p(t)= -({i}/{\hbar})[\hat H_p(t), \hat \rho_p(t)]+\mbox{$\sum_i$}\gamma_i(t)\hat {\cal L}_i \hat \rho_p(t)\,,
\label{eq1}
\ee
where $\gamma_i(t)>0$ and Lindblad terms $\hat{\cal L}_i\hat \rho_p=2\hat A_i\hat \rho_p\hat A_i^{\dag}-\{\hat A_i^{\dag}\hat A_i,\,\hat \rho_p\}$ with $\hat A_i$ being plant operators\,\cite{Gorini, Breuer3}.

\begin{figure}[b]
\includegraphics[width=0.43\textwidth, bb=0 0 206 68,clip]{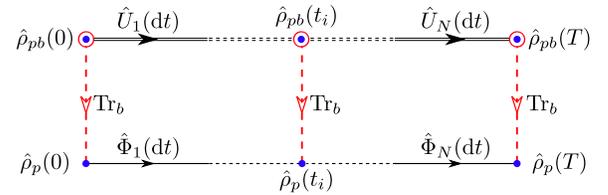}
\caption{(color online) A schematic showing how the reduced dynamics of the plant emerges from the full dynamics
of the plant-bath system by tracing over the bath state at each step.
\label{map}}
\end{figure}

{\it Reduced dynamics of the plant.}---We first discuss some general features 
of the reduced dynamics of the plant, which help understand the new criterion.
Suppose our plant-bath system evolves from $t=0$ to $t=T$. We divide this process into $N$ small segments with increment
${\rm d} t=T/N$. The entire system undergoes a unitary evolution:
\be
\hat \rho_{pb}(T)=\hat U_N({\rm d}t) \cdots \hat U_1({\rm d} t)\hat \rho_{pb}(0)\hat U^{\dag}_1({\rm d} t)\cdots \hat U_N^{\dag}({\rm d} t)\,,
\ee
where $\hat \rho_{pb}$ is the plant-bath density matrix and
$\hat U_i({\rm d}t)$ is the unitary evolution operator of the total Hamiltonian at $t_i$.
The reduced dynamics for the plant is obtained by tracing over the bath at each step,  as shown schematically in Fig.\,\ref{map}, and the density matrix of the plant evolves as:
\be
\hat \rho_p(T) = \hat \Phi_{N}({\rm d}t) \cdots \hat \Phi_{i}({\rm d}t)\cdots \hat \Phi_1({\rm d}t) \hat \rho_p(0)\,,
\ee
where $\hat \rho_p\equiv{\rm Tr}_b[\hat \rho_{pb}]$ and
the super-operator $\hat \Phi_i({\rm d}t)$ is a trace-preserving dynamical map at $t=t_i$.

In general, the dynamical map $\hat \Phi_i({\rm d}t)$
relies on the history of the plant-bath state $\hat \rho_{pb}(t_k)\; (t_k<t_i)$. For the plant to be strictly Markovian,  the bath's memory about the plant must not affect the plant's further evolution, so $\hat \Phi_i({\rm d}t)$ must depend only on the plant-bath state at $t_i$. In the simplest case, $\hat \Phi_i({\rm d}t)$
is independent of time and the state---the
dynamical map forms a semigroup with $\hat \Phi(t+\tau)=\hat \Phi(t)\hat \Phi(\tau)$.
The corresponding generator is the Lindblad super-operator $\hat {\cal L}$,
namely $\hat \Phi(t)=e^{\hat {\cal L}t}$, and the master equation for the plant is in the standard
Lindblad form: $\dot {\hat \rho}_p(t)=-(i/\hbar)[\hat H_p,\,\hat\rho_p(t)]+\hat {\cal L}\hat \rho_p(t)$.

In general, even though the dynamical map $\hat \Phi_i(dt)$ for  a non-Markovian plant depends on the history of the plant-bath system before $t_i$, for most situations this time-nonlocal dependence can be consistently accounted for by a master equation for the plant that is local in time as shown by Chru\'{s}ci\'{n}ski\,\cite{Chruscinski1}:
\begin{equation}
\dot{\hat \rho}_p(t) = \mathcal{M}_{[\hat \rho_{pb}(0)]} \hat\rho_p(t).
\end{equation}
Here the fact that $\dot{\hat \rho}_p(t)$ is written in terms of ${\hat \rho}_p(t)$, i.e., the increment in $\hat\rho_p$ only depends on $\hat\rho_p$ locally, is merely a construction. The memory effect of the non-Markovian dynamics still persists in the sense that the form of dependence ${\cal M}$ actually varies from each initial plant-bath initial state $\hat \rho_{pb}(0)$.
It is in this context that the issue of ``characterizing non-Markovianity'' is raised, where one hopes the form of $\mathcal{M}$ provides a hint on the non-Markovianity of the system by which $\mathcal{M}$ is deduced. Unfortunately, as we shall show later, there exist situations in which the {\it form} of $\mathcal{M}$ does not particularly differ from a Markovian master equation.

{\it Criterion for non-Markovianity.}---A natural way to reveal non-Markovianity is to explore the memory effect---the {\it dependence} of $\mathcal{M}$ on $\hat \rho_{pb}(0)$.  This also has an operational meaning: we should apply a time-local  unitary operation on the plant-bath system and study whether and how it influences the plant evolution at later times. Let us restrict the operation to the plant (the bath is usually uncontrollable in actual systems). In terms of dynamical maps, 
our new criterion can be phrased as:

{\it The plant's open quantum dynamics is non-Markovian if an instantaneous operation on the plant can lead to a change in its dynamical maps at later times.}

In other words, the plant dynamics is non-Markovian if $\hat \rho_{pb}(t_k)\rightarrow\hat U_p\otimes\hat I_b \,\hat \rho_{pb}(t_k) \hat U^{\dag}_p\otimes\hat I_b$ at $t_k <T$, with $\hat U_p$ a plant operator and $\hat I_b$ identity for the bath,  can lead to
\be
\hat \rho_p(T) \neq \hat \Phi_N({\rm d}t)\cdots \hat\Phi_{i}({\rm d}t)\cdots\hat\Phi_{k+1}({\rm d}t) \hat U_p \hat \rho_p(t_k)\hat U_p^{\dag},
\ee
where $\hat \Phi$ are dynamical maps before applying $\hat U_p\otimes \hat I_b$.
This new criterion, to some extent, serves as an operational definition for non-Markovianity. It
explores the memory effect in the non-Markovian dynamics---in particular, the dependence of the
initial plant-bath state. As we will show in the following example, this
criterion can indeed reveal non-Markovianity in systems of which the undisturbed evolution (before
applying the unitary operation) is indistinguishable from the Markovian one.

\begin{figure}[!t]
\includegraphics[width=0.4\textwidth, bb=0 0 299 59,clip]{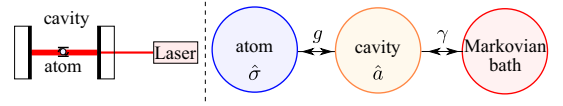}
\caption{(color online) A schematic showing the atom-cavity system. The cavity mode is coupled to the external
continuum field (a Markovian bath), and they together forms an effective non-Markovian bath for the atom. (Adapted from Ref.\,\cite{huan}).
\label{atom}}
\end{figure}

{\it Example.}---To illustrate this new criterion, here we consider the atom-cavity system as shown in
Fig.\,\ref{atom}---a two-level atom coupled to a cavity mode which in turn couples to an external
continuum---a quantum Wiener process that is equivalent to a zero-temperature Markovian bath\,\cite{Gardiner}.
If we view the cavity mode and the external continuum together as the bath, the two-level atom---the plant---is effectively
coupled to a damped cavity mode which is a non-Markovian dissipative bath, similar to the pseudo-mode model\,\cite{Imamoglu,Mazzola}.
The corresponding Hamiltonian for this system is given by\,\cite{huan}:
\begin{align}\nonumber
\hat H=&\hbar ({\omega_q}/{2})\hat \sigma_z+\hbar \, \Delta \hat a^{\dag}\hat a+\hbar\, g(\hat \sigma_-\hat a^{\dag}+\hat \sigma_+\hat a)\\&+\hbar\sqrt{\gamma}[\hat a\,\hat b_{\rm in}^{\dag}(t)+\hat a^{\dag}\,\hat b_{\rm in}(t)].
\label{eq6}
\end{align}
Here $\hat \sigma_z$ is the Pauli matrix and $\hat \sigma_{\pm}=(\hat\sigma_x\pm i\hat \sigma_y)/2$; $\hat a$ and $\hat b_{\rm in}$ are
the annihilation operators of the cavity mode and the input field of the external continuum
with $[\hat a,\,\hat a^{\dag}]=1$ and $[\hat b_{\rm in}(t),\,\hat b_{\rm in}^{\dag}(t')]=\delta(t-t')$;
$\omega_q$ is the atom transition frequency and $\Delta $ is detune frequency of the cavity mode;
$g$ and $\gamma$ are the corresponding coupling constants. After tracing over the external
continuum, the joint density matrix of the atom and cavity satisfies the following Markovian master
equation:
\begin{align}\nonumber
\dot {\hat \rho}(t)=&-i\left[ ({\omega_q}/{2})\hat \sigma_z + \Delta \hat a^{\dag}\hat a+ g(\hat \sigma_-\hat a^{\dag}+\hat \sigma_+\hat a),\,\hat \rho(t)\right] \\&+\gamma[2\hat a\hat\rho(t)\hat a^{\dag} -\{\hat a^{\dag}\hat a,\,\hat \rho(t)\} ].\label{eq3}
\end{align}
To further obtain the master equation for the atom by eliminating the cavity mode, we need to
know initial state of the atom and the cavity mode. Under the usually-applied assumption, they initially are separable
and the cavity mode is in the vacuum state: $|\psi\rangle =|\psi_a\rangle\otimes |0\rangle$. As shown
in Ref.\,\cite{huan}, the  reduced
density matrix of the atom $\hat \rho_a(t)$ satisfies a time-local master equation:
\begin{align}\nonumber
\dot {\hat \rho}_a(t)=&-i\left[\frac{\omega_q}{2}\hat \sigma_z+g\Im\{f(t)\}\, \hat \sigma_+\hat \sigma_-, \, \hat \rho_a(t)\right]\\&+g\Re\{f(t)\}\,[2\,\hat \sigma_-\hat \rho_a(t) \hat\sigma_+-\{\hat \sigma_+\hat \sigma_-,\,\hat \rho_a(t)\}].
\label{eq2}
\end{align}
Here the time-dependent function $f(t)$ satisfies the following Riccati equation:
$\dot  f(t) -i(\omega_q-\Delta+i\gamma)f(t)-g\,f^2(t)=g$ with the initial condition $f(0)=0$,
and the solution is given by\,\cite{Li}:
\be
f(t)=\frac{2g\sinh\beta t}{2\beta \cosh \beta t+(\gamma-i\,\Delta)\sinh \beta t}
\ee
with $\beta \equiv [\frac12(\gamma-i\,\omega_q+i\,\Delta)^2-g^2]^{1/2}$.
In the tuned case with $\omega_q=\Delta$ and a strong
dissipation $\gamma\ge2g$, $f(t)$ is real and positive, and we simply have:
\be\label{eq4}
\dot {\hat \rho}_a(t)=-i\left[\frac{\omega_q}{2}\hat \sigma_z, \, \hat \rho_a(t)\right]+g\,f(t)\hat {\cal L}\hat \rho_a(t)
\ee
with $\hat {\cal L}\hat \rho_a(t)=2\,\hat \sigma_-\hat \rho_a(t) \hat\sigma_+-\{\hat \sigma_+\hat \sigma_-,\,\hat \rho_a(t)\}$.
Such a master equation can also describe the case when the atom is directly coupled to
the Markovian bath but with a time-dependent coupling rate, of which the Hamiltonian is:
\be\label{eq5}
\hat H=\hbar ({\omega_q}/{2})\hat \sigma_z+\hbar \sqrt{\gamma'(t)}[\hat \sigma_-\,\hat b_{\rm in}^{\dag}(t)+\hat \sigma_+\,\hat b_{\rm in}(t)]
\ee
with $\gamma'(t)\equiv gf(t)$.
Therefore, by looking at the unperturbed evolution of the atom density matrix, we cannot tell whether the underlying dynamics is non-Markovian or not,
even though the atom-cavity interaction is highly non-Markovian when $\gamma\lesssim\omega_q$.

\begin{figure}[!b]
\includegraphics[width=0.48\textwidth, bb=0 0 477 176,clip]{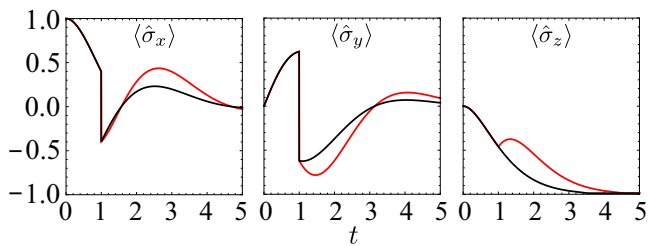}
\caption{(color online) Plot showing the time evolution of $\langle \hat \sigma_{x,y,z}\rangle$.
The black curve shows the Markovian dynamics from Eq.\,\eqref{eq12}; the red
curve shows the non-Markovian dynamics derived from the total atom-cavity evolution given in Eq.\,\eqref{eq3}.
\label{mean}}
\end{figure}

If we perturb the atom by applying a unitary operation, its density matrix evolution will deviate from
Eq.\,\eqref{eq4} due to memory of the cavity mode. To see such a deviation, the most
transparent way to look at the evolution of expectation values of the plant dynamical variables---$\langle\hat \sigma_{x}\rangle,\,\langle\hat \sigma_y\rangle,\,\langle \hat \sigma_z\rangle$, and to compare it
with their Markovain evolution which is given by [from Eq.\,\eqref{eq4}]:
\be\label{eq12}
\left[\begin{array}{c} \langle \dot{\hat \sigma}_x \rangle
\\ \langle \dot{\hat \sigma}_y \rangle\\
\langle \dot{\hat \sigma}_z \rangle \end{array}\right]=-\left[\begin{array}{ccc}
 \gamma'& \omega_q&0\\
 -\omega_q & \gamma'&0\\
 0&0&2\gamma'
 \end{array}\right]\left[\begin{array}{c} \langle \hat \sigma_x\rangle
\\ \langle \hat \sigma_y\rangle\\
 \langle \hat \sigma_z\rangle \end{array}\right]-\left[\begin{array}{c}0
\\ 0\\
2\gamma'\end{array}\right]
\ee
Such a comparison is shown in Fig.\,\ref{mean}---the initial atom-cavity state
is $(|0\rangle +|1\rangle)/\sqrt{2}\otimes|0\rangle$, $\omega_q=\Delta=g=1$, $\gamma=2g$,
and a unitary operation  on the atom: $\hat \sigma_z\otimes \hat I_c$ is applied at $t=1$.
As we can see, the Markovian and non-Markovian evolution are identical before
the unitary operation, and deviate from each other after $t=1$.

\begin{figure}[!t]
\includegraphics[width=0.45\textwidth, bb=0 0 495 172,clip]{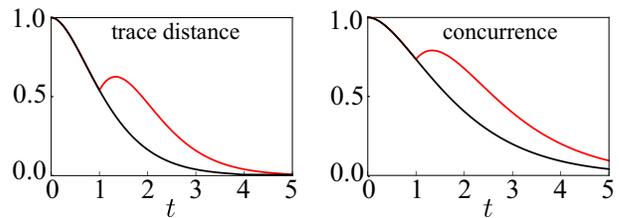}
\caption{(color online) Plot showing the time evolution of the trace distance (left) and the concurrence (right).
The red curve shows the non-Markovian evolution described by  Eq.\,\eqref{eq3} and
the black curve is the Markovian evolution from Eq.\,\eqref{eq4}.
\label{tr}}
\end{figure}

There are other figure of merits that can also quantify the deviation, e.g., the trace distance
between two different atom states\,\cite{Breuer2}, and the entanglement strength between
the atom and an introduced ancilla\,\cite{Rivas}. In the
left panel of Fig.\,\ref{tr}, we show the time evolution of the trace distance
${\rm Tr}|\hat \rho_1(t)-\hat \rho_2(t)|/2$ with two initial states for the atom
given by $\hat \rho_1(0)=\rm |1\rangle \langle 1|$ and $\hat \rho_2(0)=\rm |0\rangle \langle 0|$.
In the right panel of Fig.\,\ref{tr}, we show the time evolution of the concurrence---the ancilla entangled with the atom is another two-level system, and therefore the entanglement
strength can be quantified by the concurrence\,\cite{Hill}. Initially, the atom and the ancilla
are in the maximally-entangled state: $\frac{1}{\sqrt{2}}[|0\rangle|\otimes|0\rangle+|1\rangle|\otimes|1\rangle]$.
Other specifications are the same as those for producing Fig.\,\ref{mean}. Both the trace distance
and the entanglement strength increases after applying the local unitary operation on the atom. Such
a revival of the quantum coherence clearly indicates non-Markovianity, just as expected for such a system.

{\it Dynamical recovering.}---The increase of quantum coherence just shown, when the plant is perturbed with
local unitary operations, can be important for quantum-information processing. This allows us to recover
information of the plant that is stored in the bath, which we can call ``dynamical recovering". More importantly,
by combining local operations with the dynamical decoupling protocols---applying a sequence of control pulses\,\cite{Viloa,Biercuk,Lange}, we can characterize the plant-bath dynamics even for an unknown
bath, which can help us find the optimal strategies for maintaining the quantum coherence.

\begin{figure}[!t]
\includegraphics[width=0.45\textwidth, bb=0 0 490 175,clip]{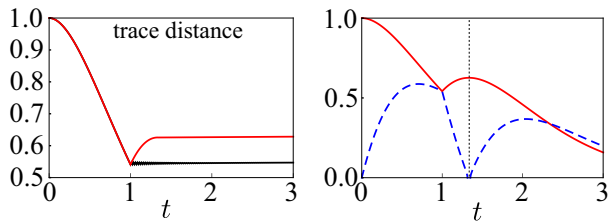}
\caption{(color online) The left panel shows the time evolution of the trace distance with a direct decoupling (black
curve) and a delayed decoupling $\Delta t=0.1$  (red curve) after applying $\hat \sigma_z\otimes\hat I_c$ at $t=1$;
the right panel shows that the maximal recovering is achieved when the atom and the cavity mode are disentangled
[red solid curve is the trace distance (same as the left panel) and the blue dashed curve is the concurrence for the atom-cavity entanglement].
\label{multi}}
\end{figure}

For illustration, we again use this atom-cavity system. In the left panel of Fig.\,\ref{multi}, we show the time
evolution of the trace distance for two different dynamical decoupling procedures: (i) the first one is a direct
decoupling at $t=1$ by applying a sequence of $\hat \sigma_z\otimes \hat I_c$ within short intervals;
(ii) the second one is a delayed decoupling (delayed by $\Delta t=0.1$) after the local operation at $t=1$.
Their difference in the trace distance tells us how much information of the atom is recovered
from the cavity mode, and it varies depending on the delay time $\Delta t$. Interestingly, the maximal difference
is achieved when the atom and the cavity mode becomes disentangled, as shown by the right panel of
Fig.\,\ref{multi}---the concurrence for the atom-cavity entanglement vanishes when
the trace distance achieves the local extremum\,\footnote{The reason why concurrence can be used to
quantify the atom-cavity entanglement is that the total occupation number is conserved
for the Hamiltonian in Eq.\,\eqref{eq6}, and the cavity mode can be treated as a two-level system.}.
This is understandable---quantum information of the atom can on longer be recovered from the cavity mode once 
they are disentangled. 

For general plant-bath interactions, starting from any moment during the evolution, there should
exist an optimal sequence of local operations on the plant for maximally recovering its information.
Operationally, one can define a measure for non-Markovianity based on such dynamical recovering.
We introduce the set of plant state pairs with the initial trace distance $D$ that is equal to 1:
$\mathcal{A}=\{\{\hat \rho_1(0),\hat \rho_2(0) \},\, \forall \, D\{\hat \rho_1(0), \hat \rho_2(0)\}=1\}$.
Suppose at moment $t$, $D\{\hat \rho_1(t),\hat \rho_2(t)\}=\alpha<1$,
the measure is:
\be
{\cal N}_{\alpha}=\max_{\forall \tau, U_{\tau}} \frac{D\{\hat \Phi_{U_{\tau}}\hat \rho_1(t),\hat \Phi_{U_{\tau}}\hat \rho_2(t)\}-\alpha}{1-\alpha},
\ee
where $\hat \Phi_{U_\tau}\hat \rho(t)\equiv {\rm Tr}_b[ \hat U_p(\tau)\otimes \hat I_b\, \hat \rho_{pb}(t)\hat U_p^{\dag}(\tau)
\otimes\hat I_b]$ is a sequence of  unitary maps for the plant from $t$ to $t+\tau$.
Obviously, for Markovian systems, $\mathcal{N}_{\alpha}$ is equal to $0$, and for non-Markovian systems in which the plant can recover its initial states via local operations, the measure is equal to $1$. In general, $\mathcal{N}_{\alpha}$ ranges between $0$ and $1$ depending on how strong the bath memory is, and also the moment $t$ that we start to apply unitary operations.

{\it Discussion.}---This criterion can be applied to study the recent experiment by Liu {\it et al.}\,\cite{Liu}. In their setup, the polarization degree of freedom of photons (the plant) couples to the frequency degrees of freedom (the bath).
During the time evolution, these two degrees of freedom become entangled---the polarization
undergoes decoherence when tracing over the frequency degrees of freedom. By changing the frequency
profile of the bath, i.e., its quantum state,
the time evolution of the trace distance for the polarization can either monotonically decrease or oscillate. This was
used to demonstrate switching between the Markovian and the non-Markovian regime. Interestingly,
if a unitary operation is applied on the polarization by flipping it at a given moment, one will find that the quantum
coherence of the polarization will revive at later times even in the so-called Markvoian regime, which indicates that
the intrinsic dynamics is non-Markovian as the frequency degrees of freedom contains memory
about the polarization. Therefore, this experiment can make a direct test of our new criterion.

{\it Conclusion.}---We have presented a new criterion for non-Markovianity in general
open quantum systems---the non-Markovianity manifests in terms of non-local
change of the dynamical map after applying a local unitary operation on the plant. It allows us
to tell whether a time-local positive map is only an artifact of a special initial quantum state
of the plant-bath system or not. We have illustrated this criterion with the atom-cavity model.
This work, on the one hand, helps clarify some subtleties of non-Markovianity in open quantum systems;
on the other hand, it provides a route to probe the non-Markovian bath and to
enhance quantum coherence in quantum-information processing.

{\it Acknowledgements.}---We thank S.L.\ Danilishin, F.Ya,\ Kahlili and other colleagues in 
the LIGO MQM group for fruitful discussions. This work has been supported by NSF grants 
PHY-0555406, PHY-0653653, PHY-0601459, PHY-0956189, PHY-1068881,
as well as the David and Barbara Groce startup fund at Caltech.


\begin{thebibliography}{99}
\bibitem{Breuer} H. P. Breuer and F. Petruccione, {\it The Theory of Open Quantum Systems}, Oxford University Press, Oxford (2007).
\bibitem{Wolf} M. M. Wolf {\it et al.}, Phys. Rev. Letter. {\bf 101}, 150402 (2008).
\bibitem{Breuer2} H. P. Breuer, E. M. Laine, and J. Piilo, Phys. Rev. Lett. {\bf 103}, 210401 (2009).
\bibitem{Rivas} A. Rivas, S. F. Huelga, and M. B. Plenio, Phys. Rev. Lett. {\bf 105}, 050403 (2010).
\bibitem{Liu} B. H. Liu {\it et al.}, Nature Phys. (Advanced online publication) (2011).
\bibitem{Viloa} L. Viola, E. Knill, and S. Lloyd, Phys. Rev. Lett. {\bf 82}, 2417 (1999);
\bibitem{Biercuk} M. J. Biercuk {\it et al.}, Nature {\bf 458}, 996 (2009).
\bibitem{Lange} G. de Lange {\it et al.}, Science {\bf 330}, 60 (2010).
\bibitem{Haikka} P. Haikka, J. D. Cresser, and S. Maniscalco, Phys. Rev. A {\bf 83}, 012112 (2011).
\bibitem{Chruscinski} D. Chru\'{s}ci\'{n}ski, A. Kossakowski, and \'{A} Rivas, Phys. Rev. A  {\bf 83}, 052128 (2011).
\bibitem{Gorini} V. Gorini, A. Kossakowski, and E. Sudarshan, J. Math. Phys. {\bf 17}, 821 (1976).
\bibitem{Breuer3} H. P. Breuer, Phys. Rev. A {\bf 70}, 012106 (2004).
\bibitem{Chruscinski1} D. Chru\'{s}ci\'{n}ski, and A. Kossakowski, Phys. Rev. Lett. {\bf 104}, 070406 (2010).
 \bibitem{Gardiner} C. W. Gardiner and P. Zoller, {\it Quantum Noise}, Springer-Verlag, Berlin,
(1991).
\bibitem{Imamoglu} A. Imamo\={g}lu,  Phys. Rev. A {\bf 50}, 3650 (1994).
\bibitem{Mazzola} L. Mazzola {\it et al.}, Phys. Rev. A {\bf 80}, 012104 (2009).
\bibitem{huan} H. Yang, H. Miao, and Y. Chen, arXiv:1108.0963 [quant-ph] (2011).
\bibitem{Li} J. Li, J. Zou, and B. Shao, Phy. Rev. A {\bf 81}, 062124 (2010).
\bibitem{Hill} S. Hill, and W. K. Wootters, Phys. Rev. Lett. {\bf 78}, 5022 (1997).

\end{thebibliography}
\end{document}